\documentclass[fleqn,usenatbib]{mnras}

\usepackage{mathptmx}

\usepackage[T1]{fontenc}

\DeclareRobustCommand{\VAN}[3]{#2}
\let\VANthebibliography\thebibliography
\def\thebibliography{\DeclareRobustCommand{\VAN}[3]{##3}\VANthebibliography}

\usepackage{subcaption}
\usepackage{graphicx}	
\usepackage{amsmath}	
\usepackage{amssymb}	
\usepackage{tikz}
\usepackage{blindtext}
\usepackage{placeins}
\usepackage{xcolor}

\title[A real-time AGDP at ORT]{A real-time Automated Glitch Detection Pipeline at Ooty Radio Telescope}

\author[J. Singha et al.]{
Jaikhomba Singha ,$^{1}$\thanks{E-mail : mjaikhomba@gmail.com}
Avishek Basu,$^{4,2}$
M. A. Krishnakumar,$^{3}$
\newauthor
Bhal Chandra Joshi,$^{2}$\thanks{E-mail: bcj@ncra.tifr.res.in}
~P. Arumugam$^{1}$
\\
$^{1}$Department of Physics, Indian Institute of Technology Roorkee, Roorkee-247667, India\\
$^{2}$National Centre for Radio Astrophysics, Tata Institute of Fundamental Research, Pune 411007, India\\
$^{3}$Fakult{\"a}t f{\"u}r Physik, Universit{\"a}t Bielefeld, Postfach 100131, D-33501 Bielefeld, Germany\\
$^{4}$Jodrell Bank Centre for Astrophysics, University of Manchester, Oxford Road, Manchester, M13 9PL, UK
}

\date{Accepted XXX. Received YYY; in original form ZZZ}

\pubyear{2020}

\begin{document}
\label{firstpage}
\pagerange{\pageref{firstpage}--\pageref{lastpage}}
\maketitle

\begin{abstract}
 Glitches are the observational manifestations of superfluidity inside neutron stars. The aim of this paper is to describe 
 an automated glitch detection pipeline,  
 which can alert the observers on possible real-time detection of rotational glitches in  pulsars. Post alert, the pulsars can be monitored at a higher cadence to measure the post-glitch recovery phase. Two algorithms namely, Median Absolute Deviation (MAD) and polynomial regression have been explored to detect glitches in real time. The pipeline has been optimized with the help of simulated timing residuals for both the algorithms. Based on the simulations, we conclude that the polynomial regression algorithm is significantly more effective for real time glitch detection. The pipeline has been tested on a few published glitches. This pipeline is presently implemented at the Ooty Radio Telescope. In the era of upcoming large telescopes like SKA, several hundreds of pulsars will be observed regularly and 
such a tool will be useful for both real-time detection as well as optimal
utilization of observation time for such glitching pulsars.
\end{abstract}




\section{Introduction}  \label{sec:intro}
    Pulsars are highly magnetized, rapidly rotating neutron stars with remarkably stable rotational period.  Despite this stability in their rotation rate, they exhibit two types of timing irregularities: glitches and timing noise. Glitches are the abrupt changes in the rotation rates of pulsars~\citep{Radman69} whereas timing noise refers to the slow wander in their rotation rates~\citep{Boynton_1972,Cordes_Helfand1980, Tempo2II, Lyne+2013, parthasarthy_etal_2019}. Glitches are rare events usually observed in young pulsars. However, till date two millisecond pulsars have shown small glitches \citep{Cognard2004, McKee2016}. The initial model of pulsar glitch considered the spin-up to be explained by considering the change in the moment of inertia of the star by sudden relaxation of neutron star crust from its stressed oblate structure. Such a model, referred to as ``\textit{starquake model}'' \citep{starquakeruderman1969} could not explain the occurrence of the second Vela glitch \citep{Reichley+Downs+1971}. The most accepted model till date is based on the pinned superfluid vortices inside the neutron star crust \citep{AndersonItoh1975,AlparvortexcreepII1984,AlparvortexcreepI1984}. With temperature typically of order $10^6 - 10^8$ K inside the neutron star \citep{Sauls1989}, neutrons are expected to be in superfluid state. The condensate's wavefunction does not support macroscopic rotation, hence superfluid of neutrons breaks up into an array of quantized vortices to mimic the bulk rotation. The energetics of the superfluid vortices lead to the pinning of the vortex core to the crustal inhomogeneities. The process of pinning results in freezing-out of the superfluid's velocity, whereas the outer crust slows down due to the external spin-down torque. Such a scenario generates a stress between the superfluid component and the outer crust, which when finally released is observed as a pulsar glitch. The superfluid reservoir's moment of inertia can be estimated from the  observed spin-up. The evidence of large glitches hints towards the neutron star core's contribution towards the pulsar glitch \citep{Montoli+2020, Pizzochero+2020, Basu_2018} along with the contribution from the crust. \\
    \indent A remarkable feature of pulsar glitches is the post-glitch exponential recovery of rotational frequency. The characteristic timescale of recovery helps in probing the microphysics governing the interaction between the normal and the superfluid components \citep{Sauls1989, recovery1, recovery2}.  The process of re-coupling of the decoupled superfluid component governs such post-glitch behaviour. The relaxation period is varied in nature and could last for days or even months \citep{espinoza2011,Yu2013}. It is important to point out that short recoveries (relaxation time scales $<$ 10 days) have been reported only for a small fraction of glitching pulsars. This could partly be due to the low cadence of observations (2-4 weeks) of most glitching pulsars \citep{Yu2013}. Hence, a more complete picture of post-glitch recovery behaviour with small relaxation time scales can be studied only through high cadence extended timing after a glitch. Therefore, an ability to detect glitches in real time is required along with an observing strategy with good control over the cadence of observation after a glitch. \\ 
    \indent In this paper, we present the methodology adopted to design a real-time automated glitch detection pipeline (AGDP) which can provide alerts (in the form of emails) on the possible detection of a glitch. The underlying algorithm of AGDP is sensitive to the systematic pattern due to timing noise and is capable of detecting glitches with a wide range of glitch amplitudes. The structure of the pipeline has been discussed in Section \ref{sec:pipeline}. In Section \ref{chrz}, we have presented methods to simulate the Time of Arrival (ToA) residuals by injecting artificial glitches. These simulated residuals were further used to tune the pipeline and obtain the optimal parameters for reliable detections by AGDP. Finally,  we present the detection of glitches in real data  for pulsars, observed with the Ooty Radio Telescope \citep{swarup1971large} using AGDP in Section \ref{sec:testgl} and end the paper with conclusion and discussions in Section \ref{sec:conc}.
	\begin{figure}
\includegraphics[scale =0.4]{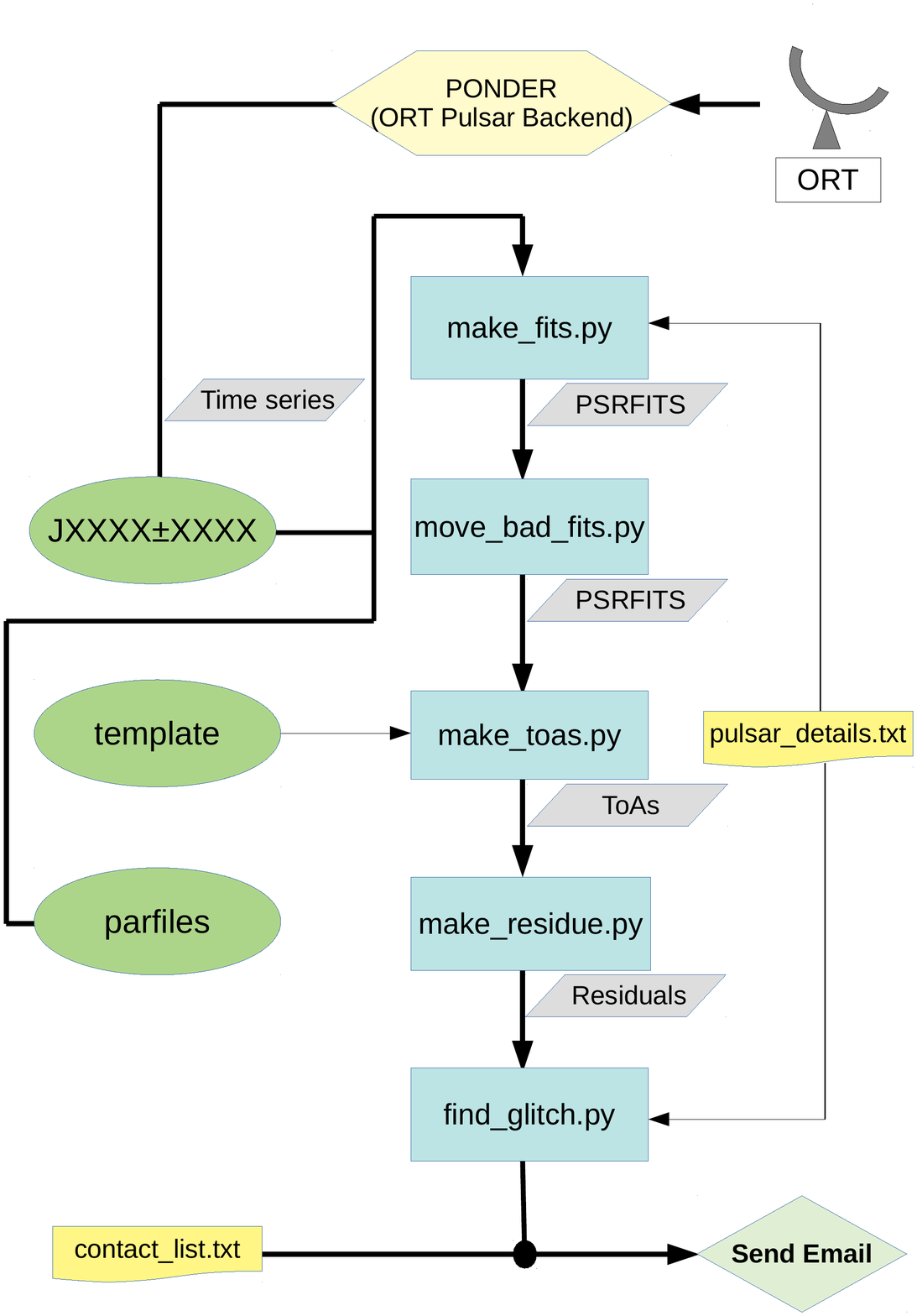}
\caption{Diagrammatic representation of the functioning of the real-time AGDP, implemented at the Ooty Radio Telescope (ORT). PONDER is the ORT pulsar backend \citep{Naidu_2015}. The rectangular blocks represent all the python codes used in the pipeline. These codes are executed one after another as indicated by the arrows. The pulsar directories (JXXXX$\pm$XXXX) and directories with the pulsar templates (\texttt{template}) are shown in oval blocks. The updated pulsar ephemeris are located inside the directory \texttt{parfiles}.} The ASCII file \texttt{pulsar\_details.txt} contains the details  required by the pipeline (observation duration, frequency of observation, dispersion measure, bandwidth, optimum window size $w$, threshold $T$, the start MJD, etc) for all the pulsars to be analysed. The email ids of the users to  whom the glitch alarm needs to be sent are given in \texttt{contact\_list.txt}. The parallelogram blocks represent the input/output at different stages of the pipeline.
\label{block}
\end{figure}
\section{Description of the pipeline} \label{sec:pipeline}
The main aim of AGDP is to give an alarm whenever there is a possibility of the presence of a glitch in the timing residuals in real time. In order to do this, AGDP performs statistical analysis on the timing residuals to determine if the pulsar has glitched or not, based on some user defined statistical cut-off. Before the pipeline performs such statistical analysis on the residuals, it requires the data to pass through multiple stages of analysis. In this section, we have discussed these stages of analysis that are performed by this pipeline. In Figure \ref{block}, a schematic diagram of the functioning of the pipeline is presented.
\begin{table*}
\begin{tabular}{|c|c|}
\hline
\hline
\textbf{Name of the directory} & \textbf{Constituents } \\
\hline
\texttt{codes}   &  All python codes, \texttt{pulsarfile.txt} and \texttt{contacts.txt} \\
JXXXX$\pm$XXXX & Two sub-directories for time series files and PSRFits : \texttt{Tims} and \texttt{Fits}\\
\texttt{template} & Noise free pulsar templates \\
\texttt{residuals} & Timing residuals of all the pulsars \\
\texttt{parfiles} & Pulsar parfiles with updated solutions \\
\hline
\end{tabular}
    \caption{Description of the various directories used in AGDP. JXXXX$\pm$XXXX are the individual pulsar directories in their J2000 names.  }

\label{dir}
\end{table*}
\subsection{Structure and functional description of the pipeline}
\label{struct}
AGDP is a python based pipeline dependent on several pulsar data analysis softwares like \texttt{DSPSR}\footnote{\url{http://dspsr.sourceforge.net/}} \citep{DSPSR}, \texttt{PSRCHIVE}\footnote{\url{http://psrchive.sourceforge.net}} \citep{psrchive2, PSRCHIVE} and \texttt{TEMPO2}\footnote{\url{https://bitbucket.org/psrsoft/tempo2/src/master/}} \citep{tempo2hobbsetal, Tempo2II}. This pipeline has five basic programs (a) \texttt{make\_fits.py}, (b) \texttt{move\_bad\_fits.py}, (c) \texttt{make\_toas.py}, (d) \texttt{make\_residue.py} and (e) \texttt{find\_glitch.py}. The  pipeline follows a well defined  nomenclature for various directories.  All pulsar names are specified as J2000 names. The first step in the pipeline is to generate PSRFITS \citep{psrchive2} files using DSPSR by folding the input time series files to produce the required number of sub-integrations for every epoch of observation (\texttt{make\_fits.py}). An updated pulsar ephemeris provided in the folder \texttt{parfiles} is used for this purpose. In the next step, the PSRFITS files with signal to noise (S/N) > 20 are filtered out for further analysis (\texttt{move\_bad\_fits.py}). The Time of Arrival (ToAs) of the pulses are then obtained by cross-correlating a template (given in the directory, \texttt{template}) with the observed epoch's folded profile using \textit{pat} in \texttt{PSRCHIVE} thus obtaining the required number of ToAs per epoch (\texttt{make\_toa.py}). Further, using \texttt{TEMPO2}, the post-fit residuals of the timing solution are generated (\texttt{make\_residue.py}).  Finally, some statistical tests on the ToA residuals are performed (\texttt{find\_glitch.py}) in order to detect outliers. An alarm is generated on detection of a pulsar glitch. Majority of the pulsars which glitch also exhibit strong timing noise, so the ToA residuals exhibit a systematic pattern. Two different techniques to search for glitches in the timing residuals, in the form of an outlier, have been employed in \texttt{find\_glitch.py}. To account for the systematic pattern in the residuals, a window is chosen consisting of a few fixed number of points in the residuals a window of w days/observation epochs, consisting of 3$\times$w data points (assuming 3 integrations per epoch) is chosen. Then, the window is made to slide over observation epochs in forward direction till the latest epoch's residual. If the threshold criteria is met, an email is sent to the observers with their email ids listed in the text file \texttt{contact\_list.txt}. The pipeline is run for all pulsars given in a particular list (\texttt{pulsar\_details.txt}), which also contains all the required parameters (observation duration, frequency of observation, dispersion measure, bandwidth, optimum window size $w$, threshold $T$, the start MJD, etc). The various directories used by the pipeline along with their constituents have been listed in Table \ref{dir}. 
 \begin{figure*}
\includegraphics[width=17.5cm,height=10cm]{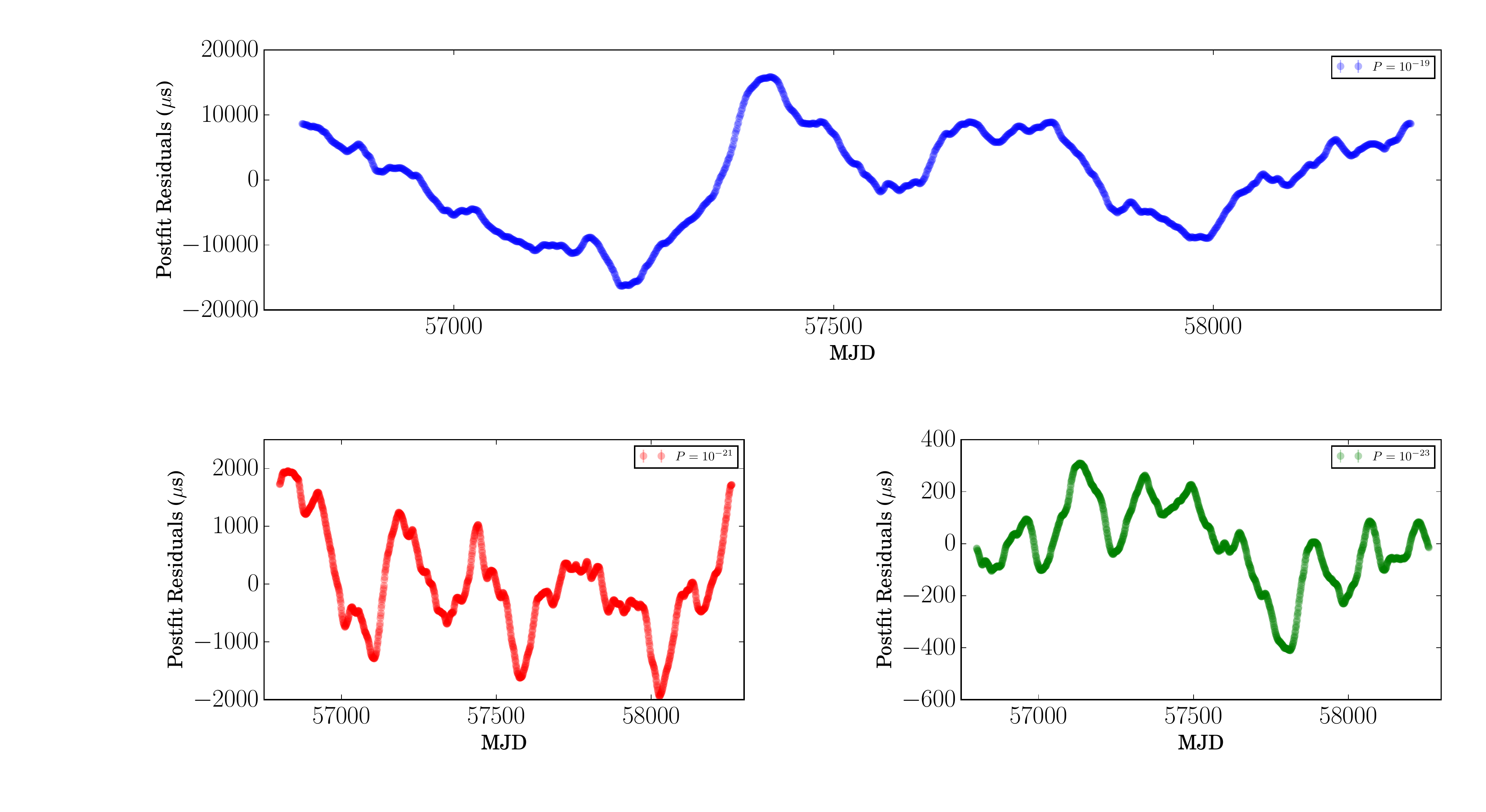}
\caption{Three out of the nine combinations of  simulated timing noise residuals using the \texttt{simRedNoise} plugin of \texttt{TEMPO2} are shown for amplitudes  of $10^{-19}, \, 10^{-21}$ and $10^{-23}$ respectively as indicated in the inset. The cut-off frequency is  $f_c$ 1.0 $yr^{-1}$ and the index is chosen as 4.0}
\label{simrednoise}
\end{figure*}
\subsection{Glitch detection algorithms} \label{glitch_det_Algo}
In this section, we present the statistical methods that were adopted for detecting the pulsar glitches. Glitches cause a pulsar to spin-up. This means that the model no longer tracks the data and the ToAs arrive earlier and earlier (compared to the model) which leads to a ramp-like feature in the residuals. So we adopt statistical measures which are capable of identifying the post-glitch ToA residuals as outliers from the pre-glitch residuals.
The median absolute deviation (MAD) statistics is known to be a robust outlier detector \citep{MADleys}. Therefore, we have chosen the MAD statistics in order to detect glitches. However, the presence of systematic pattern due to timing noise pose a serious challenge and can be addressed in a better way using the linear regression analysis with the polynomial features ~\citep{poly}. A short description of how these two statistical methods have been applied to the pulsar timing data has been presented below.
\subsubsection{Median Absolute Deviation}
The median absolute deviation (MAD) is a robust estimator for an outlier.The MAD is computed over a fixed number of points in the residuals covered by a window of chosen number of days. Then, the window is made to slide by one observation epoch at a time. If the window contain less number of samples, the estimates of MAD and the median of samples will show statistical fluctuations, which decreases with the increase in the number of sample points. We account for these statistical fluctuations in defining our threshold for identifying an outlier. The criteria for detection of pulsar glitch using MAD algorithm is 
\begin{equation}
X_i \geq T(M_w + \sigma_{w,M}) +(m_w + \sigma_{w,m})
\end{equation}
where $X_i$ is the post-fit residual of  $i^{th}$ ToA, $T$ is the threshold for detection, $M_w$
is the MAD associated with  window "$w$" including the $i^{th}$ point, $\sigma_{w,M}$ is the corrective factor associated with the measurement of MAD due to under sampling, $m_w$ is the median computed for the same number of points and $\sigma_{w,m}$, the corrective factor associated with the measurement of median for a given window size. By using a window length with few number of ToA residuals, the estimates of MAD and median become insensitive to the timing noise. Hence, $\sigma_{w,M}$ and $\sigma_{w,m}$ have been estimated from white noise ToA residuals spanning over 3 yrs. We compute both MAD and median for a fixed window "$w$" over the full data span. Such analysis results in generating $x = L/l$ number of MAD and median estimates. Here, $L$ is the number of ToA residuals over three years and $l$ is the number of ToAs in the given window length. $\sigma_{w,M}$ and $\sigma_{w,m}$ were estimated by calculating the standard deviation over $x$ number of estimates. This process was repeated for all choices of window length.
\subsubsection{Linear regression with polynomial features}
Polynomial regression (linear regression with polynomial features) is a simple regression technique with the traditional least-squares loss function. In order to take care of the stochastic wander caused by the presence of timing noise in the residuals, polynomial features of degree three \citep{pulsartiming} have been utilised in the regression method. As in the case for MAD, we apply the regression method over a window, which keeps sliding in order to span the entire data. For a window size $w$ in days, we first obtain the number of data points based on the number of sub-integrations and the cadence of observations. Let us assume this number to be $N_w$. The first $N_w-3$ residuals are used to train the regression model and the final 3 residuals are predicted using the model. We then compare the predicted and the observed residuals to conclude if the points are outliers. In order to do this comparison, the standard deviation of the regression model applied to the training set, $\sigma_{N_w-3}$ is obtained. Further, we set the detection criteria as the following, 
\begin{equation}
X_{j\{P\}} > T_{w}\sqrt{\sigma_{N_w-3}^2 + \sigma_j^2}
\end{equation}
where $X_{j\{P\}}$ is the difference of the predicted and observed ToA residuals for any of the last three data points ($j$), $T_w$ is the threshold factor for detection and $\sigma_j$ is the ToA error of the $j^{th}$ ToA. It has to be noted that this condition needs to be satisfied for all three data points to flag these data points as outliers and therefore, flag a glitch. We have selected three residuals for this criteria since in our case, we are using 3 sub-integrated profiles. We apply this regression method using the functions available in the python library, sklearn.linear\_model\footnote{\url{https://scikit-learn.org/stable/modules/generated/sklearn.linear_model.LinearRegression.html}}$^,$\footnote{\url{https://scikit-learn.org/stable/modules/generated/sklearn.preprocessing.PolynomialFeatures.html}}.
 
\section{Characterization by Simulations and Optimization}
\label{chrz}
\begin{figure}
\centering
\includegraphics[scale =0.5]{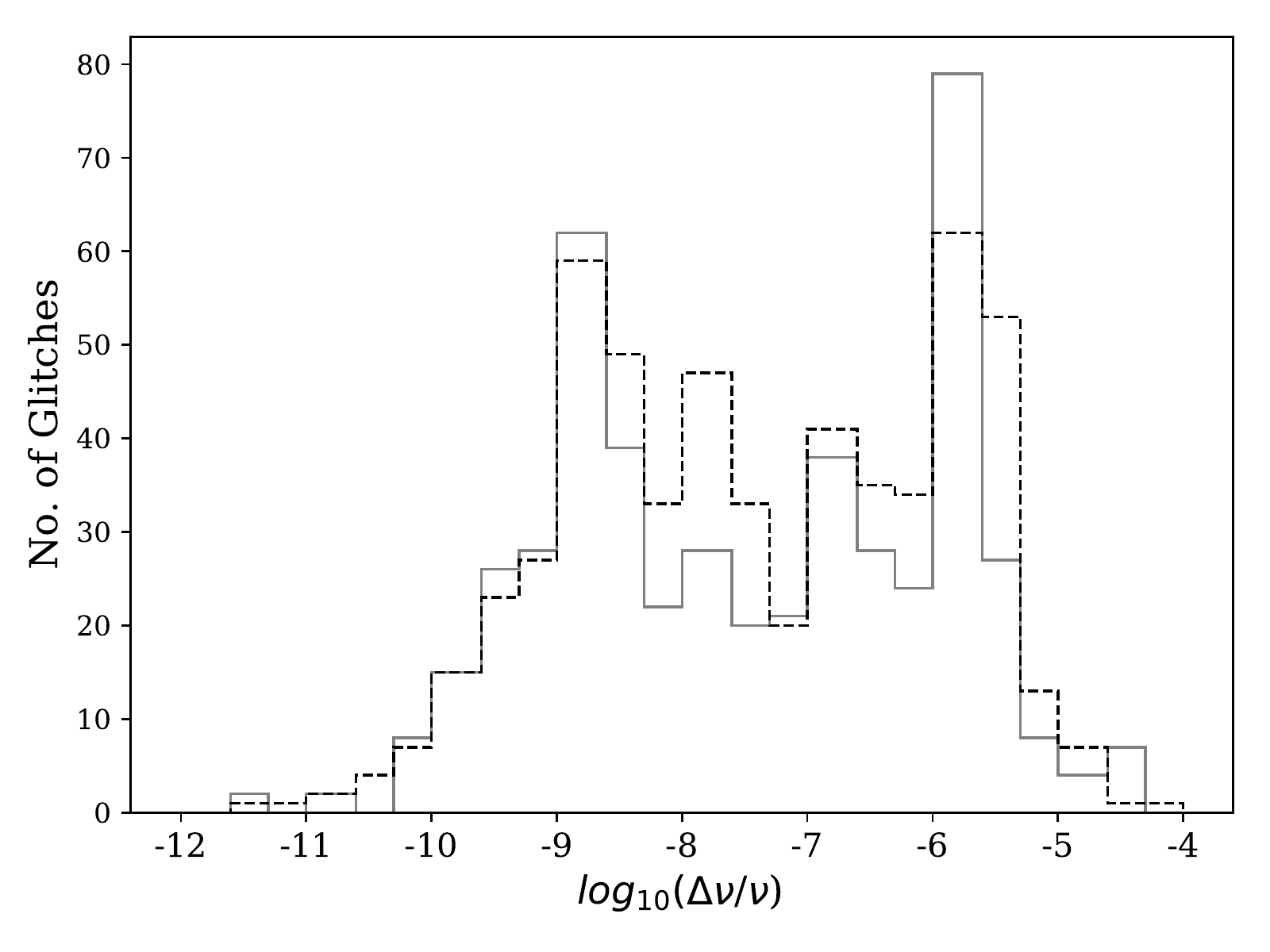}
 
\caption{ The distribution of $log_{10}(\frac{\Delta \nu}{\nu})$ is shown above. The black color dashed histogram is obtained from the real data whereas the grey color histogram is constructed using Smirnov inverse transformation. It shows that the simulated distribution preserves the bi-modality in the distribution. The real data has been taken from the pulsar glitch catalogue available at the Jodrell Bank website. A total of 568 glitches in 192 pulsars with reported measurements of $(\frac{\Delta \nu}{\nu})$ have been detected as of January 1, 2021.}
\label{cdfgl}
\end{figure}

In order to maximize the number of true detections and minimize the number of false alarm, it is essential to obtain an optimal choice  for $T$ and $w$. Since, different pulsars exhibit different strength of timing noise and different glitch amplitudes, it is expected that both $T$ and $w$ will have dependence on the timing noise parameters and the glitch amplitudes. Such optimal choice for $T$ and $w$ can be obtained by running our detection algorithm on various realization of timing residuals spanning the full range of glitch distribution shown in Figure \ref{cdfgl} and different timing noise parameters. However, the scarcity of real glitch events, spanning this range, makes our sample size small enough to obtain  robust estimates of optimal $T$ and $w$. Therefore, we have simulated the timing residuals with different timing noise parameters and glitch amplitudes. In the following sections, we have presented the methodology adopted to simulate the timing residuals with glitches and timing noise. 

\subsection{Step I: Simulating timing noise}
\label{step1}
Timing noise in pulsars is a slow wandering of the ToA residuals obtained after subtracting a simple spin down model of the pulsar. Such variation indicates non secular evolution of the rotation frequency, which is 
stochastic in nature. This wandering is intrinsic to the pulsar  and is independent of any propagation effects. The typical time scale of such wander can be of the order of a year to few years \citep{Lyne+2013}. The power spectrum density of timing noise can be written as ~\citep{lasky2015},
\begin{equation}
\label{density}
\mathcal{P}(f) = A\, \Big (1 + \frac{f^2}{f_c^2}\Big)^{-\frac{\alpha}{2}}
\end{equation}
The quantity $A$ is the amplitude of the power spectral density, $\alpha$ is the index and $f_c$ is the spectral turn around frequency in the units of year $^{-1}$. Integrating Equation \ref{density} over all frequencies $( 0,\infty)$ gives the total power $P$ as
\begin{equation}
\label{totalpower}
P = \int_{0}^{\infty} \mathcal{P}(f) df = \sqrt{\frac{\pi}{4}} A f_c \frac{\Gamma(\frac{\alpha -1 }{2})}{\Gamma(\frac{\alpha}{2})} 
\end{equation}
Equation \ref{totalpower} is valid for $\alpha$ greater than 1. We have used the \texttt{fake} plugin of \texttt{TEMPO2} to  simulate the ToAs with white noise residuals. The time series with timing noise residuals were constructed from the white noise residuals using the \texttt{simRedNoise} plugin of \texttt{TEMPO2}. These time series were simulated assuming a daily cadence with 3 sub-integrations per observation. We have constructed 9 such realization of ToA residuals with different combination of the timing noise parameters $A$ and $f_c$ keeping $\alpha = 4$. Such a choice of timing noise parameters have been made in order to identify small glitches even in presence of strong timing noise. In Figure \ref{simrednoise}, we have shown three representative simulated timing noise residuals. 

\begin{figure*}
\begin{center}
\includegraphics[scale = 1.0]{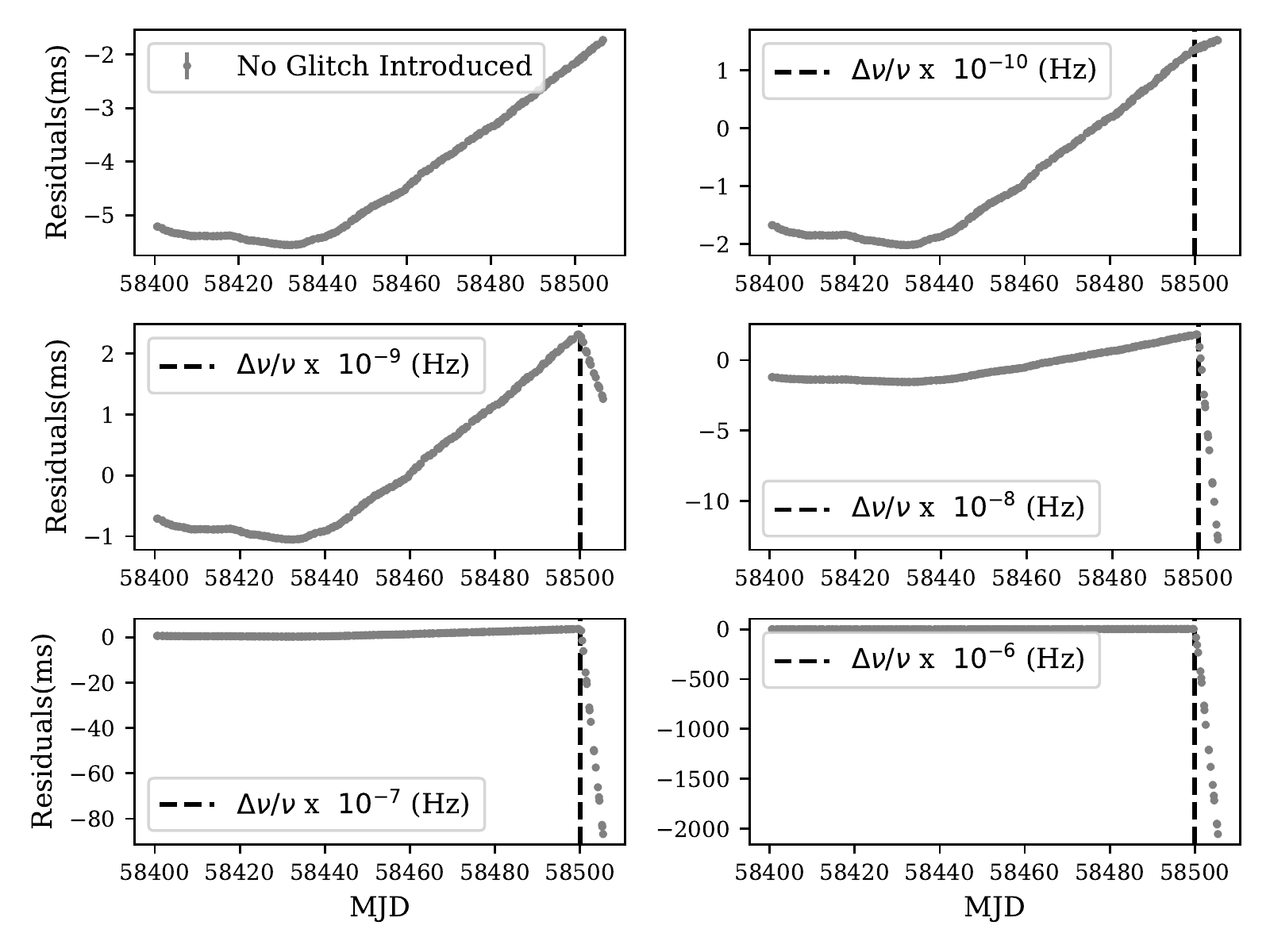}
\caption{Figure shows the residuals for a few selected cases of glitch amplitudes ranging from $\frac{\Delta \nu}{\nu} = 10^{-10}$ to $\frac{\Delta \nu}{\nu} = 10^{-6}$. Evidently, very small glitch $\frac{\Delta \nu}{\nu} \sim 10^{-10}$ is indistinguishable from the timing noise for $A = 10^{-19}$ and $f_c = 0.50$  $yr^{-1}$. The effect of the spin up in the phase connected solution appears distinctively
from the glitch size greater than $\frac{\Delta \nu}{\nu} = 10^{-8}$. AGDP was successful in detecting 4 out of 5 glitches. The glitch with $\frac{\Delta \nu}{\nu} = 10^{-10}$ was not detected for this specific simulation.}
\label{simglt}
\end{center}
\end{figure*}
\subsection{Step II: Simulating glitch parameter}
\label{step2}
We have used $\frac{\Delta \nu}{\nu}$ and $\frac{\Delta \dot{\nu}}{\dot{\nu}}$\footnote{$\Delta \nu$ and $\Delta \dot{\nu}$ is defined as $(\nu_{post-glitch} - \nu_{pre-glitch})$ and $(\dot{\nu}_{post-glitch} - \dot{\nu}_{pre-glitch}$) respectively.} from the pulsar glitch catalog available at the Jodrell Bank website \footnote{\url{http://www.jb.man.ac.uk/pulsar/glitches/gTable.html}}. As of January 1, 2021, 568 glitches in 192 pulsars have been detected, with relative increase
    in the spin frequency $\frac{\Delta \nu}{\nu}$ ranging between $ 10^{-12}$ to $10^{-4}$. The Pearson correlation test statistics between $\frac{\Delta \nu}{\nu}$ and $\frac{\Delta \dot{\nu}}{\dot{\nu}}$, was 0.07, implying no correlation between them. Hence both of them were independently re-sampled from their distribution by the method of inverse transformation \citep{devroye1986}. The cumulative distribution function (CDF) was created from the data obtained from the Jodrell Bank glitch catalogue. The $\frac{\Delta \nu}{\nu}$ and $\frac{\Delta \dot{\nu}}{\dot{\nu}}$ were interpolated as the function of CDF. The CDF ranges between 0 and 1, hence ``$N$'' random numbers were generated from a uniform distribution bounded between 0 and 1, which were used to obtain the simulated values of the $\frac{\Delta \nu}{\nu}$ and $\frac{\Delta \dot{\nu}}{\dot{\nu}}$ from the interpolation function. 
Figure \ref{cdfgl} shows the distribution of the $\log_{10} \frac{\Delta \nu}{\nu}$. The black color dashed line represents the actual distribution from the real data and the grey color distribution obtained from inverse transformation. It is clear that the method of inverse transformation can mimic the true bi-modality in the distribution. We have generated 400 (N=400) events from our simulation. 
\begin{figure*}
\centering
\caption{The maps of  $P(D)$, $P(\bar{D})$, $P(f)$ and $l-1$ norm for $A = 10^{-19}$ and $f_c = 0.25$ ($yr^{-1}$) for the two different algorithms are presented in this figure. $P(D)$ is the fraction of true detection of glitches, $P(\bar{D})$ is the fraction of non-detection of glitches and $P(f)$ is the fraction of false alarm. Similar plots were produced for  nine sets of timing noise parameters for each of two algorithms used.}
\centering
\includegraphics[width=16cm,height=11cm]{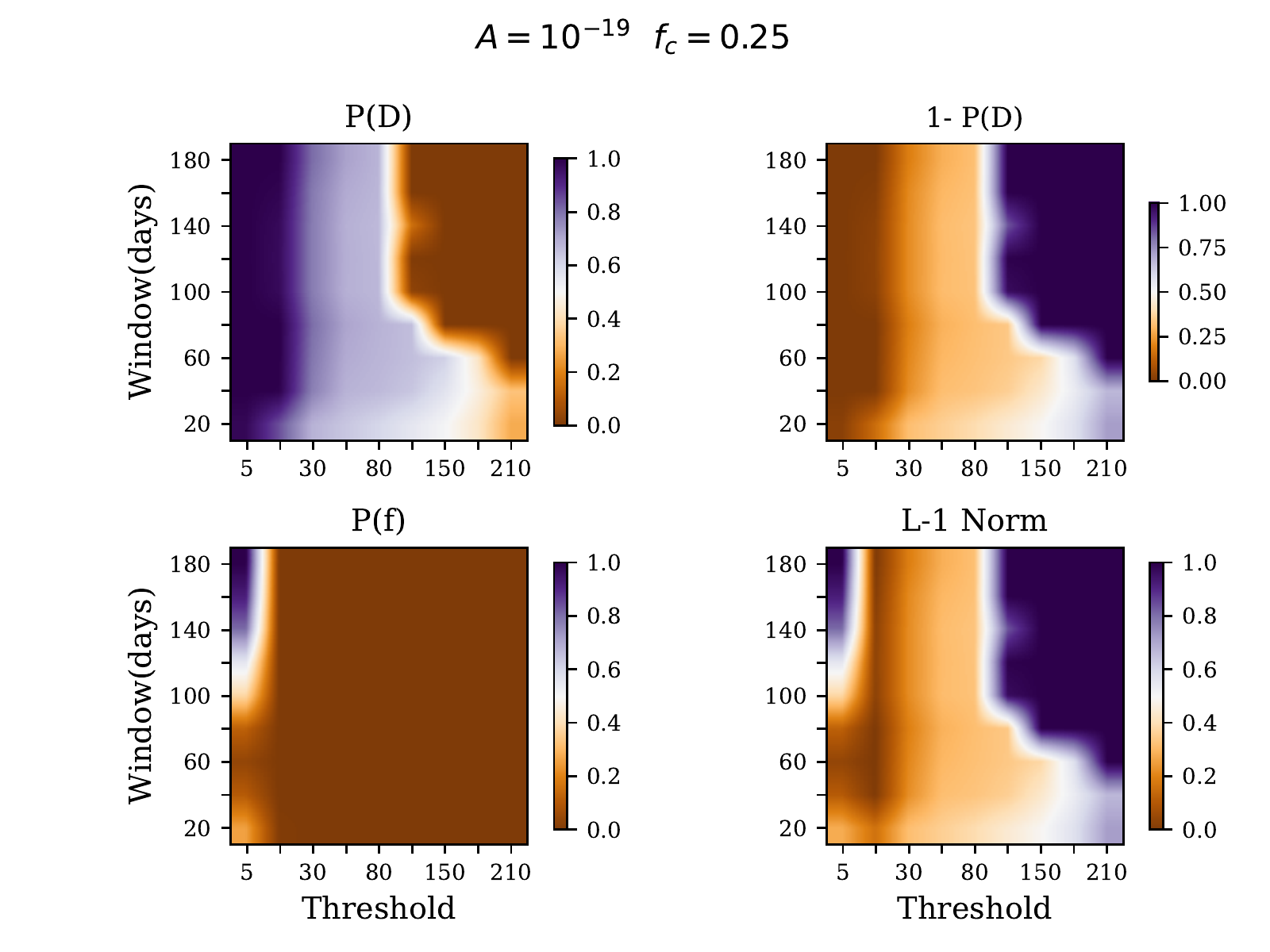}

\subcaption[]{Maps using Median Absolute Deviation Algorithm}
\label{maps1}
\includegraphics[width=16cm,height=11.4cm]{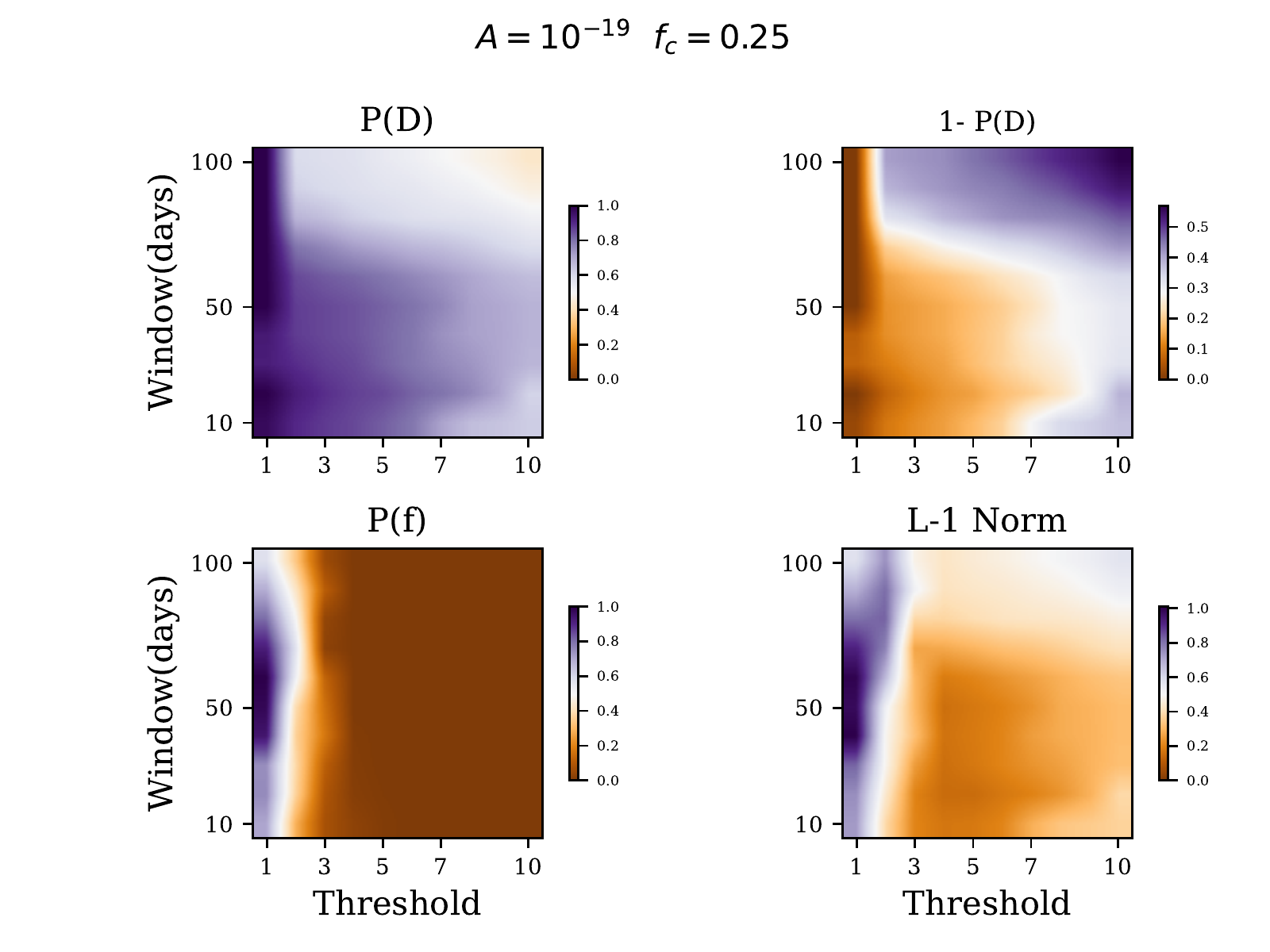}
\subcaption[]{Maps using Linear Regression with Polynomial Features}
\label{maps2}
\end{figure*}
\begin{table*}
\caption{The results of the optimisation with respect to the optimum window size, $w$ (days), and threshold, $T$ are presented in this table. $A$ and $f_c$ ($yr^{-1}$) are the timing noise parameters. $P(D)$, $P(\Bar{D})$ and $P(f)$ are the fraction of glitches detected, the fraction of non-detection and the fraction of false alarm respectively. For different sets of timing noise parameters we present the fraction of true glitch detection and the false alarm for the two algorithms (Median Absolute Deviation and Polynomial Regression) used in AGDP.}
\centering
\begin{tabular}{cccccccccc}
\hline
\hline
\multicolumn{2}{c}{Timing Noise} & \multicolumn{4}{c}{Median Absolute}  & \multicolumn{4}{c}{Linear Regression} \\
\multicolumn{2}{c}{Parameters} & \multicolumn{4}{c}{Deviation}  & \multicolumn{4}{c}{with polynomial features} \\
\hline 
A  & $f_c$ ($yr^{-1}$) & w(days)  & T & P(D) & P(f) & w(days) & T & P(D) & P(f) \\
\hline
\centering
$10^{-19}$   & 0.25  & 60 & 10 & 0.725 & 0 & 20 & 4 & 0.89 & 0.015      \\
\hline
$10^{-19}$  & 0.50  & 160 & 10 & 0.6275 & 0.2808  & 10 & 4 & 0.86 & 0.018 \\
\hline
$10^{-19}$  & 1.0  & 20 & 10 & 0.155 & 0.371 & 100 & 1 & 1 & 0.008 \\
\hline
$10^{-21}$  & 0.25  & 100 & 10 & 0.78 & 0  & 10 & 4 & 0.96 & 0 \\
\hline
$10^{-21}$  & 0.50  & 120 & 10 & 0.68 & 0  & 10 & 5 & 1 & 0 \\
\hline
$10^{-21}$  & 1.0  & 180 & 10 & 0.555 & 0  & 10 & 5 & 1 & 0.012  \\
\hline
$10^{-23}$  & 0.25  & 80 & 10 & 0.78 & 0 & 20 & 3 & 0.99 & 0 \\
\hline
$10^{-23}$  & 0.50  & 160 & 10 & 0.83 & 0  & 20 & 3 & 0.99 & 0 \\
\hline
$10^{-23}$  & 1.0  & 80 & 10 & 0.78 & 0  & 80 & 3 & 1 & 0 \\
\hline
\hline
\end{tabular}
\label{opttab}
\end{table*}
\begin{figure*}
\includegraphics[scale=0.85]{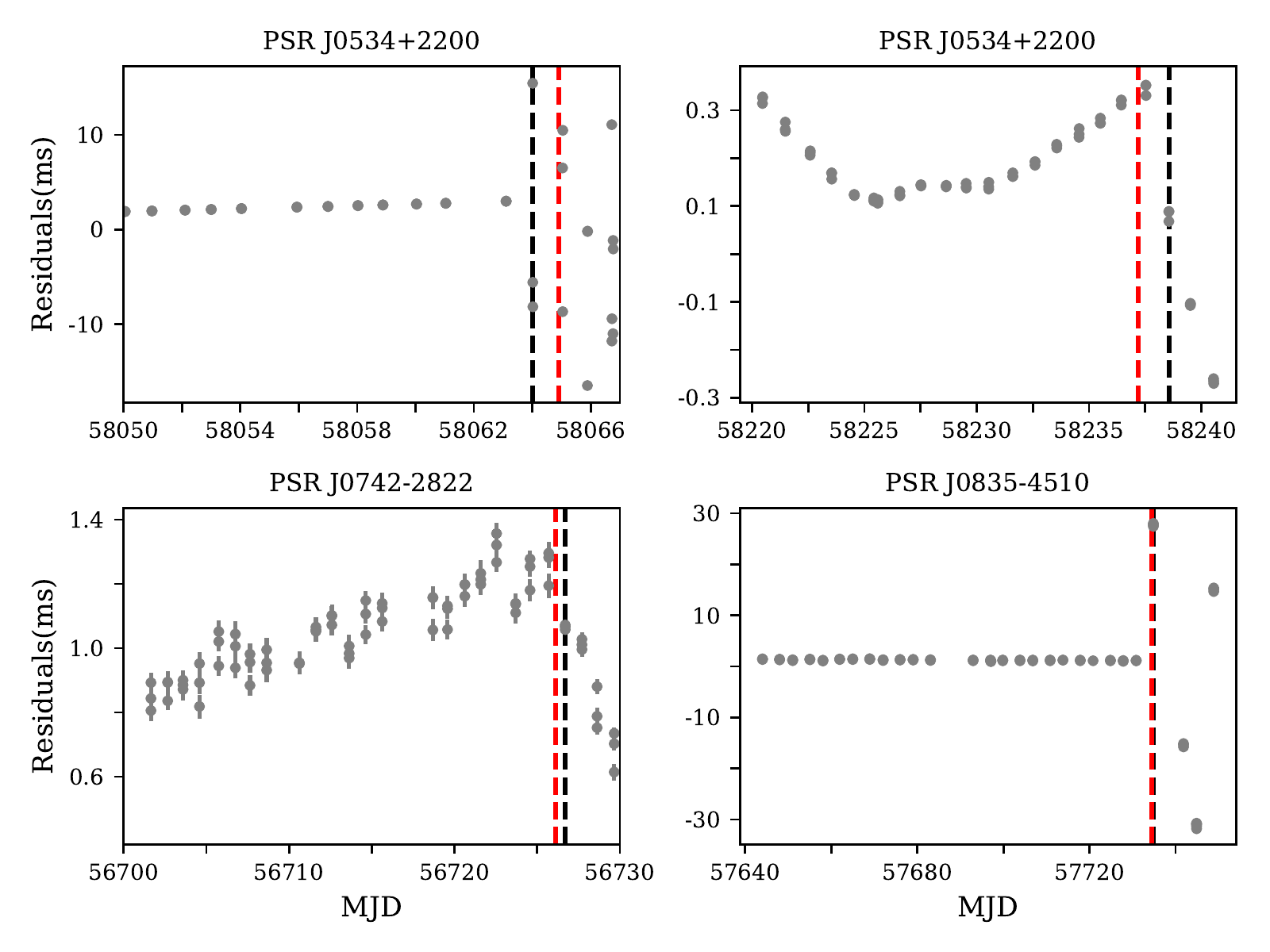}
\caption{The timing residuals for recently published glitches  ~\citep{basu2020}, which were used for testing AGDP are shown in these plots. The red dashed lines depict the published glitch epochs given in Table \ref{glrltab}.  The black-dashed lines are the epochs where AGDP gives an alarm of probable glitch detection. The glitches, which occurred on MJD 58064.9(1) in J0534+2200 and MJD 57734.4(2) in J0835-4510, are large glitches. On the other hand, the glitches that occurred on MJD 58237.2(1) in J0534+2200 and MJD 56726.1(2) in J0742-2322 are small glitches.  Since the residuals after the glitch have not been fitted, for small glitches, there is a small ramp seen in the residual plots after the glitch, but after large glitches, the residuals randomly move in all directions because of the loss in phase connections. }
\label{gltresi}
\end{figure*}

\subsection{Step III: Simulating glitched ToAs}
If the rotation model of the star is known with good precision, ToAs can be predicted by computing the rotation phase ($\phi$) from the prior knowledge of the model parameters. The phase is obtained by integrating the Taylor series expansion of the rotation frequency ~\citep{Tempo2II},
\begin{equation}
\label{taylor}
\phi(t) = \sum_{n\geq 1} \frac{\nu ^{(n-1)}(t_0)}{n!} (t - t_0)^n + \phi_0
\end{equation}
In the above equation the quantity \{$t_0$,$\nu ^{(n)}(t_0)$\} is the ephemeris of the pulsar and $\phi_0$ is some reference phase, which can be assumed to be zero without any loss of generality. Putting $\phi_0 = 0$ implies that the counting of pulsar's rotation starts from $t = t_0$. However, the sudden occurrence of a glitch at any particular epoch $t_g$ (called as glitch epoch) in pulsars leads to an abrupt change in the rotational parameters. Such changes incorporate a pattern in the timing residuals, where the ToA residuals are observed to show a ramp like feature in the negative direction as shown in the Figure \ref{simglt}. The ramp is explained from the unaccounted change in the rotation phase $\phi_g$ because of the changed rotational parameters after a glitch. $\phi_g$  can be expressed as \citep{tempo2hobbsetal, Tempo2II, Yu2013} 
\begin{equation}
\label{phig}
\begin{array}{cc}
\phi_g = \Bigg \{ & 
\begin{array}{cc}
\Delta \phi + \Delta \nu_p(t-t_g) + \frac{1}{2} \Delta \dot{\nu}_p(t-t_g)^2\\+ \{1 - e^{-(\frac{t-t_g}{\tau})}\} \Delta \nu_d \tau & t \geq t_g \\ \\
0 &  t < t_g. \\
\end{array}
\end{array}
\end{equation}
The quantity  $\Delta \nu_{p}$ and $\Delta \dot{\nu}_p$  are permanent changes in the rotational frequency and it’s first derivative, $\Delta \nu_{d}$ is the part of change in glitch which decays exponentially with some time-scale, $\tau$. So the total change in rotational frequency and it’s first derivative at the glitch epoch are given by,  $\Delta \nu_{g}$ = $\Delta \nu_{p}$ + $\Delta \nu_{d}$ and $\Delta \dot{\nu}_{g}$ = $\Delta \dot{\nu}_{p}$ + $\frac{\Delta \nu_{d}}{\tau}$ respectively. The quantity  $\Delta \phi $ is added to account for any discontinuity in phase. $\Delta \nu_p$ has linear contribution in phase. Since the pipeline always searches for glitches locally within a defined window length, the contribution from this term is co-variant with the first order term in the exponential recovery. Hence, we ignore the effects of $\Delta \nu_p$. For simplicity, we also assume total recovery in the glitch to its pre-glitch rotational frequecny and neglect $\Delta \phi$ and $\Delta \dot{\nu}_p$ 
by setting them to zero. 
With these assumptions, Equation \ref{phig} can be simplified as:
\begin{equation}
\label{phigr}
\begin{array}{cc}
\phi_g = \Bigg \{ & 
\begin{array}{cc}
\{1 - e^{-(\frac{t-t_g}{\tau})}\} \Delta \nu_d \tau & t \geq t_g \\ \\
0 &  t < t_g. \\
\end{array}
\end{array}
\end{equation}
A particular epoch have been arbitrarily chosen as the glitch epoch. For any future epochs after $t=t_g$, the phase change due to glitch have been computed using Equation \ref{phigr}.
Since the pulsars spin up during glitches, the measured ToA ($t_m$) is smaller than the predicted ToA ($t_p$). Hence, the difference $\delta t$ between them can be written as $t_p -t_m = \delta t$. In our case, the ToA obtained with the timing noise is equivalent to $t_p$, and $\delta t = \frac{\phi_g(t_p)}{\nu(t_p)}$. Therefore, the ToAs for the simulated glitches (equivalent to $t_m$) can be obtained uisng $t_m = t_p -\delta t$. The variation in the glitch amplitude has been incorporated in $\phi_g$ via $\Delta \nu_d$. We use the value of $\nu_0$, $\dot{\nu}_0$ and $t_0$ of the Crab pulsar given in the ATNF pulsar catalogue \citep{ATNFcatalogue}. The choice of Crab pulsar is arbitrary. A typical value of $\tau$= 30 days have been chosen. Using the value of $\nu(t_g)$, we compute $\Delta \nu_d =(\frac{\Delta \nu}{\nu})_s \nu(t_g)$. We performed this simulation with 400 glitch parameters mentioned in Section \ref{step2} and using 9 time series containing the timing noise parameters discussed in Section \ref{step1}. In Figure \ref{simglt}, we present the phase connected solution of the simulated glitched ToAs. It can be seen that a very small glitch of size $\frac{\Delta \nu}{\nu} \, \sim 10^{-10}$ is almost indistinguishable from the timing noise of the pulsar. The effect of spin up in the phase connected solution becomes clearly visible for the ones with glitch sizes greater than or equal to $10^{-8}$. Amongst the specific simulations for glitches shown in Figure \ref{simglt}, only the glitches with amplitude greater than or equal to $10^{-9}$ were detected. However, in some instances of our simulations even glitches with amplitudes $10^{-10}$ are detected. 

\subsection{Step IV: Optimization}
\label{step4}
In order to obtain an optimal choice of $T$ and $w$ for all the 3600 time series, we run our detection algorithm to search for the epoch where the glitches were artificially injected. To study the dependence of $T$ and $w$ on the strength of the timing noise and the glitch amplitude, we perform this search of artificially injected 400 glitches for every timing noise parameter set separately. We construct a 2D grid of $T$ and $w$. For every pair of $T$ and $w$, we compute the following three quantities to characterize their dependence on $T$ and $w$. 
\begin{enumerate}
    \item The fraction of glitches detected $P(D)$.\newline
    $P(D)$ is defined as the number of glitches detected out of $N$ (in our case $N=400$) number of glitches injected. Hence $P(D)=\frac{D}{N}$. \newline 
    \item The fraction of non-detection $P(\Bar{D})$.\newline 
    $P(\Bar{D}) = 1-\frac{D}{N}$ \newline
    \item The fraction of false alarm $P(f)$.\newline
    The presence of strong timing noise can sometimes be misinterpreted as a glitch by the detection algorithm for some specific combination of $T$ and $w$. We define $P(f) = F/\bar{N}$. Here $F$ is the number of times the false alarm gets generated for every pair of $T$ and $w$. Whereas $\bar{N}$ is the maximum number of false alarm generated after running the detection algorithm over all pairs spanning the grid, for a particular case of timing noise parameter. $P(f)$ is an important parameter which need to be minimized.
\end{enumerate}
Since both $P(\bar{D})$ and $P(f)$ are quantities less than unity, we construct
the $l-1$ norm constructed out of the $P(\bar{D})$ and$P(f)$. The $l-1$ norm is defined as,
\begin{equation}
\Phi_{l-1}(w,T) = P(\bar{D}) + P(f).
\end{equation}
We find the particular combination of $T$ and $w$ for which the $l-1$ norm is minimum. As this choice maximises the glitch detection rate while minimising the false alarms, we consider it to be the optimal choice. As an illustration, the maps corresponding to $A = 10^{-19}$ and $f_c = 0.25$ for the two different algorithms used, are shown in Figures \ref{maps1} and \ref{maps2}. We extract the optimised $T$ and $w$ from the $l-1$ maps.  The optimal choice of $T$ and $w$ for different timing noise parameters assuming a daily cadence and 3 sub-integrations per epoch is presented in Table \ref{opttab}. We also carried out simulations with cadence of 10 days and 3 sub-integrations per observation. The conclusions from the analysis of lower cadence simulations suggested a longer window size and higher threshold. For instance, the optimal window size and threshold for polynomial regression in the case of $A=10^{-19}$ and $f_c=0.25$, for a cadence of 10 days, becomes $w = 30$ days and $T=5$. Although the results of these low cadence simulations are not presented in the paper, our procedure is general enough to simulate the results for any choice of cadence.

\begin{table}
\caption{This table lists the glitch parameters for all the real glitches used in this work. The first column is the J name of the pulsars, followed by the epoch of the glitch and the fractional change in rotation frequency~\citep{basu2020}.}
\centering
\begin{tabular}{cccc}
\hline
\hline
Pulsar  & Glitch Epoch &  $\frac{\Delta \nu}{\nu}$ \\
J Name & MJD & $\times 10^{-9}$ \\
\hline
\centering
J0534+2200 & 58064.9(1) & 484.39(1) \\
J0534+2200 & 58237.2(1) & 1.7(6) \\
J0742-2322 & 56726.1(2) & 2.6(2) \\
J0835-4510 & 57734.4(2) & 1433.2(9) \\
\hline
\hline
\end{tabular}
\label{glrltab}

\end{table}

\section{Testing AGDP on real residuals} \label{sec:testgl}
AGDP was tested on the timing residuals of four hand picked glitches recently published by \cite{basu2020} using the ORT and uGMRT. The glitch epoch and the fractional spin-ups are given in Table \ref{glrltab}. Firstly, with the help of spectral modelling using \texttt{TEMPO2}~\citep{tempo2hobbsetal, spectralmod, spectralmod2}, the timing noise parameters were determined for each of the pulsars for which the real data was used. The timing noise parameters for the Crab pulsar are obtained as $A \sim 10^{-19}$, $f_c \sim 0.25$ and $\alpha \sim 3.6$. The timing noise parameters of other pulsars and their details will be provided in a second paper on timing irregularities (J. Singha et al, in preparation).
The timing noise parameters obtained were used to determine the relevant $T$ and $w$ from Table \ref{opttab}. It is seen from Table \ref{opttab} that for polynomial regression, the true glitch detection rates are higher compared to the MAD algorithm. Hence, only the parameters obtained for polynomial regression have been used. AGDP, with this choice of parameters, was applied on the real data. The AGDP algorithm was capable of detecting all the four glitches. These detections are shown in Figure \ref{gltresi}. Since AGDP has been designed to run in real time, phase wraps after the glitch epoch are not added, which leads to a scatter in the ToA residuals after the glitch epoch. For all these residuals, the epoch where the pipeline gives an alarm(in the form of emails) is indicated with black-dashed lines and the real glitch epochs are indicated by red-dashed lines in Figure \ref{gltresi}.
\section{Conclusions and Future Plans} \label{sec:conc}
    We have successfully developed a real-time automated glitch detection pipeline (AGDP) for reducing the raw pulsar data to produce the ToA residuals and performing statistical tests on these residuals to search for glitches. We have also run simulations in order to obtain the optimum $T$ and $w$ for different timing noise models. Using these values of $T$ and $w$, the pipeline has been optimized. To demonstrate the ability of this pipeline to detect glitches, we tested it on four real glitches offline. The procedure developed in this paper can be used to determine the optimum window size and threshold with the data simulated for any desired choice of cadence. This will allow AGDP to be useful for any glitching programme.\\
\indent AGDP has been already implemented at the Ooty Radio Telescope(ORT) for the last two months. The future telescope, like the Square Kilometre Array(SKA), will observe several pulsars regularly. Therefore, AGDP with its capability to detect glitches in real time will serve as a very useful tool to devise an optimal observing strategy for studying short duration post-glitch recoveries. 

For future development, we plan to extend the pipeline to include other possibilities like finding automatically the glitch epoch and the relative spin up during a glitch.

\section*{Acknowledgements}
We would like to thank all the staff members of the Radio Astronomy Center and all the operators of the ORT who have assisted during the observations. The ORT is operated and maintained at the Radio Astronomy Centre by the National Centre for Radio Astrophysics of the Tata Institute of Fundamental Research. BCJ and AB acknowledge the support of the Department of Atomic Energy, Government of India, under project \# 12-R\&D-TFR-5.02-0700.  AB also acknowledges the support from the UK Science and Technology Facilities Council (STFC). Pulsar research at Jodrell Bank Centre for Astrophysics and Jodrell Bank Observatory is supported by a consolidated grant from STFC.

\section*{Data Availability}
The data underlying this article will be shared on reasonable request to the corresponding author. All the codes are available in the following link : \url{https://github.com/mjaikhomba/AGDP}. 



\bibliographystyle{mnras}
\bibliography{reference} 




\appendix


\bsp	
\label{lastpage}
\end{document}